\documentclass[showpacs,twocolumn]{revtex4}
\usepackage{epsfig}

\begin{document}

\noindent {\bf Watanabe and Hu Reply:} In a recent Letter
\cite{04prl}, Watanabe, Yukawa, Ito, and Hu (WYIH) proposed the
idea of superscaling for percolation on rectangular domains and
used numerical data near the effective critical density
$\rho_{c}'$ to support such idea. In \cite{Comment}, Pruessner and
Moloney (PM) pointed out some problems in \cite{04prl}. Most of
these problems are caused by the differences in using $\rho_{c}'$
or the bulk critical density $\rho_{c}$ in the analysis. Figure 2
of \cite{04prl} shows clearly that at $\rho=\rho_c$ the existence
probabilities $E_p$ for different aspect ratios $R$ are different
\cite{Cardy}. Thus it is natural that we consider a horizontal
line at $E_p=0.5$ in Fig. 2(a) to define the effective critical
density $\rho_{c}'$ for curves with different aspect ratios $R$
and using data near $\rho_{c}'$ to study superscaling shown in
Fig. 4 of \cite{04prl}. Please note that at $\rho_c'$, the
correlation length is of the order of vertical linear dimension
$L$. By considering $E_p$ near $\rho_{c}'$ rather than near
$\rho_{c}$, we will not have the problems mentioned at the second
and the third paragraphs of \cite{Comment}. To avoid the
confusion, in the abstract and the first paragraph of \cite{04prl}
we should replace ``deviation from the critical point $\epsilon$''
by ``deviation from the effective critical point $\rho_c'$,
$\epsilon$,''; we should also replace the term ``the critical
point'' below Eq. (1) and ``$\rho_c$'' of Eq. (2) by ``the
effective critical point $\rho_c'$'' and ``$\rho_c'$'',
respectively.

The value of $b$ in Eq. (10) of \cite{04prl} was determined by PM
to be $3/8 (\equiv y_t/2)$ based on
$P(L,R,\epsilon)= P(RL,R^{-1},\epsilon)$, while WYIH obtained $b =
0.05$. This discrepancy is caused by differences in definitions of
percolating clusters. PM used only the largest cluster to define
the percolation probability $P$ and thus obtained
$P(L,R,\epsilon)=P(RL,R^{-1},\epsilon)$. WYIH used all clusters
extending from the top to the bottom of the system to define
$P$ and have $P(L,R,\epsilon) \ge
P(RL,R^{-1},\epsilon)$ for $R > 1$. Besides, WYIH used the
effective critical density $\rho_c'$ in Fig. 7 of \cite{04prl} to
determine $b=0.05$, while PM did not used $\rho_c'$. To avoid the
confusion, we should also use the effective critical density
rather than the critical density to define $t$ in Eq. (10) of
\cite{04prl} as in the case of Fig. 7.


Since a horizontal line at $E_p=E_p(L,0,1)$ in Fig. 2 of
\cite{04prl} has been used to defined $\rho_c'$, Fig. 2 indicates
that $\rho_c'$ depends on $L$ and $R$ and $\lim_{L \rightarrow
\infty} E_p(L,\rho_c'-\rho_c,R) - E_p(L,0,R)\neq 0$ for $R\neq 1$.
Thus PM correctly pointed out that the difference between using
$\rho_c'$ and $\rho_c$ does not vanish in the limit $L \to \infty$
for $R\neq 1$. For this reason, we should replace ``correct for
finite-size effects'' above Eq. (9) in \cite{04prl} by ``study
superscaling behavior''.


\begin{figure}[htb]
\epsfig{file=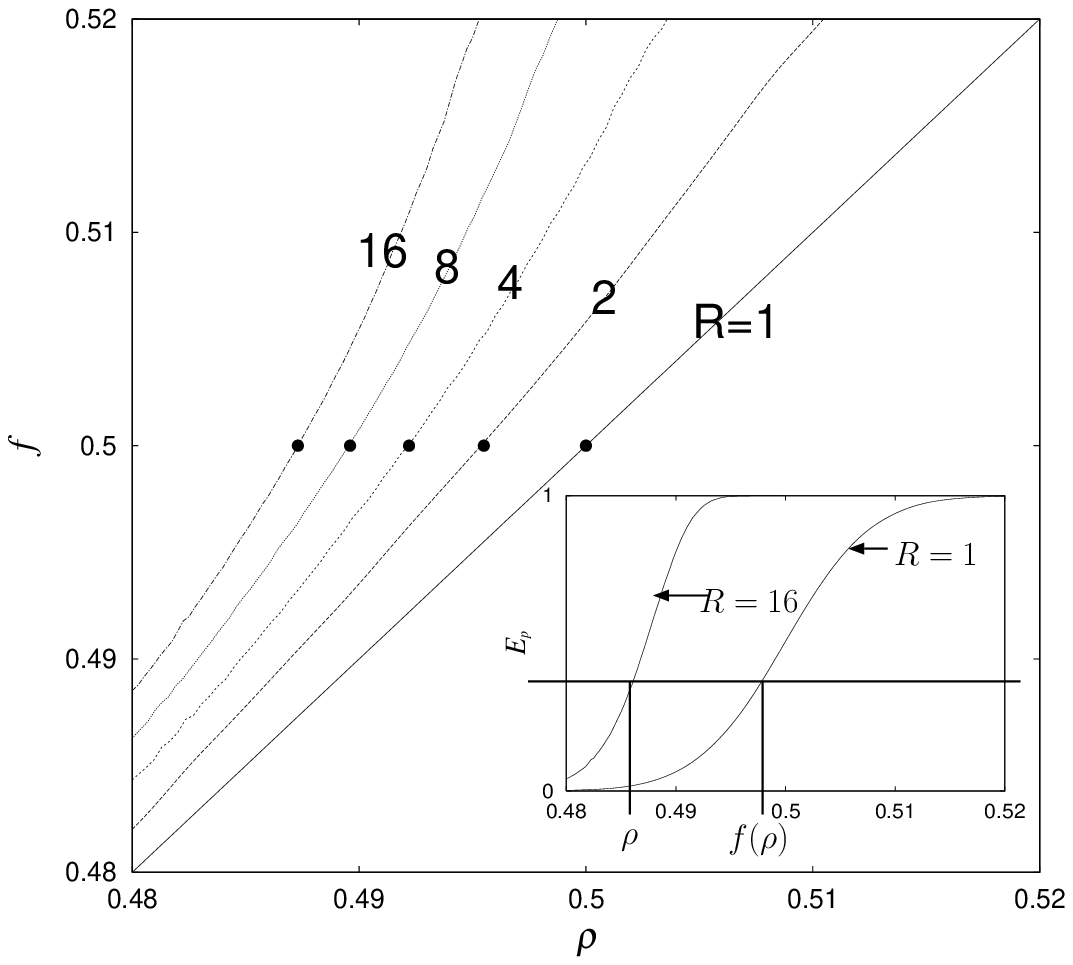,width=8cm}
\caption{
The mapping functions $f$ in Eq.~(1), which are
nonlinear for $R > 1$. The inset shows how to obtain $f$ from $E_p$.
The solid circles are the effective critical points $\rho_c'$.
}
\label{fig_f}
\end{figure}

The meaning of the {\it nonlinear scaling variable}
questioned by PM can be understood
from the mapping function $f(\rho)$ defined by
$E_p(L,\rho,R) = E_p(L,f(\rho),1)$,
with $f$ determined numerically and plotted in Fig. 1 which shows
nonlinear dependence on $\rho$ for $R > 1$ and is consistent with
Fig. 6 and Table I of \cite{04prl}. Finite-size scaling implies
$E_p(L,f(\rho),1) \sim F((f(\rho)-\rho_c)L^{y_t})$; combining this
with the expansion of $f(\rho)$ at $\rho_{c}'$ up to the
linear term with $({\partial f}/{\partial \rho})_{\rho_c'} \sim
R^a$ gives
$E_p(L,\rho,R) \sim F((\rho-\rho_{c}')L^{y_t}R^a)$,
which shows superscaling of $E_p$ with scaling variable
$x=(\rho-\rho_{c}^{'})L^{y_t}R^a$ and $\rho_{c}'$
(instead of $\rho_c$) enters $x$ naturally.
We find that
$\ln ({\partial f}/{\partial \rho})_{\rho_c'}$ increases linearly
with $\ln R$ with a well defined exponent $a=0.14$ \cite{05wh} which is
consistent with that obtained in \cite{04prl}.
Similar calculations of the mapping functions of the
percolation probability give the exponent $b=0.05$ \cite{05wh}
which is also consistent with that in \cite{04prl}.






{\it Scaling} can be considered as ``data collapse'' for critical
systems with different sizes \cite{stanley99}. We found very good
data collapse for systems with different aspect ratios in
\cite{04prl}. It is appropriate to call such results
``superscaling''.

This work was supported by Grants
NSC 93-2112-M 001-027 and NSC 94-2119-M-002-001.
\vskip 2 mm
\noindent
Hiroshi Watanabe$^{1}$ and Chin-Kun Hu$^2$

$^1$Department of Complex Systems Science,
Graduate School of Information Science, Nagoya
University, Furouchou, Chikusa-ku, Nagoya 464-8601, Japan

$^2$Institute of Physics, Academia
Sinica, Nankang, Taipei 11529, Taiwan

\vskip 2 mm
\noindent
Received 10 June 2005\\
PACS numbers 05.50.+q, 24.10.Lx, 89.75.Da

\end{document}